\documentclass[sigconf,nonacm]{acmart}

\usepackage{times}  
\usepackage{helvet}  
\usepackage{courier}  
\usepackage[]{url}  
\usepackage{graphicx} 
\usepackage{natbib}  
\usepackage{caption} 
\usepackage[ruled,vlined]{algorithm2e}

\usepackage{booktabs}
\usepackage{multirow}
\usepackage{comment}
\usepackage{amsmath}
\usepackage[export]{adjustbox}
\usepackage{xcolor}

\newcommand{\etal}{{\em{et al.}}}

\AtBeginDocument{%
  \providecommand\BibTeX{{%
    \normalfont B\kern-0.5em{\scshape i\kern-0.25em b}\kern-0.8em\TeX}}}

\setcopyright{acmcopyright}
\copyrightyear{2022}
\acmYear{2022}
\acmDOI{10.1145/1122445.1122456}

\begin{document}

\title{Meta-Shop: Improving Item Advertisement For  Small Businesses}

\author{Yang Shi}
\affiliation{%
  \institution{Rakuten, Inc.}
  \country{San Mateo, USA}
}
\email{yang.shi@rakuten.com}

\author{Guannan Liang}
\affiliation{%
  \institution{Rakuten, Inc.}
  \country{San Mateo, USA}
}
\email{guannan.liang@rakuten.com}

\author{Young-joo Chung}
\affiliation{%
  \institution{Rakuten, Inc.}
  \country{San Mateo, USA}
}
\email{youngjoo.chung@rakuten.com}

\renewcommand{\shortauthors}{Shi and Liang and Chung}

\begin{abstract}
In this paper, we study item advertisements for small businesses. This application recommends prospective customers to specific items requested by businesses. From analysis, we found that the existing Recommender Systems (RS) were ineffective for small/new businesses with a few sales history. Training samples in RS can be highly biased toward popular businesses with sufficient sales and can decrease advertising performance for small businesses. We propose a meta-learning-based RS to improve advertising performance for small/new businesses and shops: \textit{Meta-Shop}. 
Meta-Shop leverages an advanced meta-learning optimization framework and builds a model for a shop-level recommendation. It also integrates and transfers knowledge between large and small shops, consequently learning better features in small shops.
We conducted experiments on a real-world E-commerce dataset and a public benchmark dataset. Meta-Shop outperformed a production baseline and the state-of-the-art RS models. Specifically, it achieved up to 16.6\% relative improvement of Recall@1M and 40.4\% relative improvement of nDCG@3 for user recommendations to new shops compared to the other RS models.
\end{abstract}

\ccsdesc[500]{Information systems~Recommender systems}
\ccsdesc[300]{Information systems~Similarity measures}

\keywords{cold-start recommendation, recommender system, meta-learning}


\maketitle

\section{Introduction}
Recommender systems (RS) are the core component of the E-commerce  business. E-commerce companies such as Amazon and eBay need to provide personalized recommendations to customers and provide prospective user targeting services to sellers, i.e., recommend potential customers to sellers so that sellers can increase their sales on the platform. 
When E-commerce companies provide such services to sellers or business owners, one obstacle is the recommendation bias to large businesses. While large and established businesses with a sufficient number of popular items can get accurate potential customers, the performance of RS for small businesses is usually unsatisfactory due to items with a limited number of sales. 
In addition, the bias toward items in large businesses could hurt users' experience since the system may fail to find new and niche items in small businesses, which may be of interest to the users. 



We show the different distribution of sales, revenue, and unique items of large and small businesses from Rakuten Ichiba in Figure~\ref{fig:1}. Rakuten Ichiba\footnote{https://www.rakuten.co.jp} is the largest E-commerce platform in Japan, where 50K shops sell more than 300M items. Each shop sells various products, from unique hand-made shoes to home appliances. Shops also promote their products via targeted advertisement and various customer experiences such as gift wrapping services or rewards.
  While items from the small businesses account for 20\% of all items, their revenue contribution is only 6\%. With a similar average price per item in the big and small businesses (\$29 and \$35 respectively), we conclude that small businesses are not finding the right customers. By improving recommendation performance for small businesses, we aim to increase the revenue contribution from small businesses and the satisfaction from both sellers and customers.


\begin{figure}
\begin{minipage}{\linewidth}
\includegraphics[width = 0.7\linewidth,center]{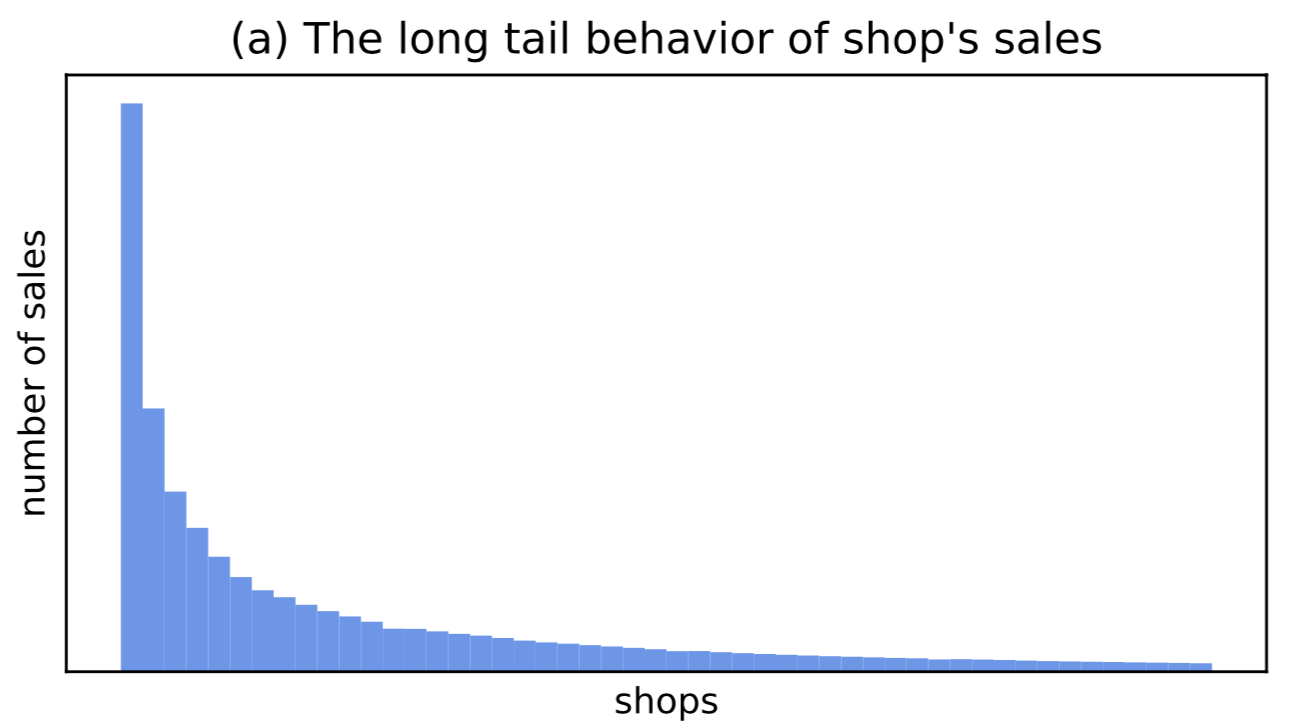}
\end{minipage}
\begin{minipage}{\linewidth}
\centering
   \includegraphics[width = 0.7\linewidth,center]{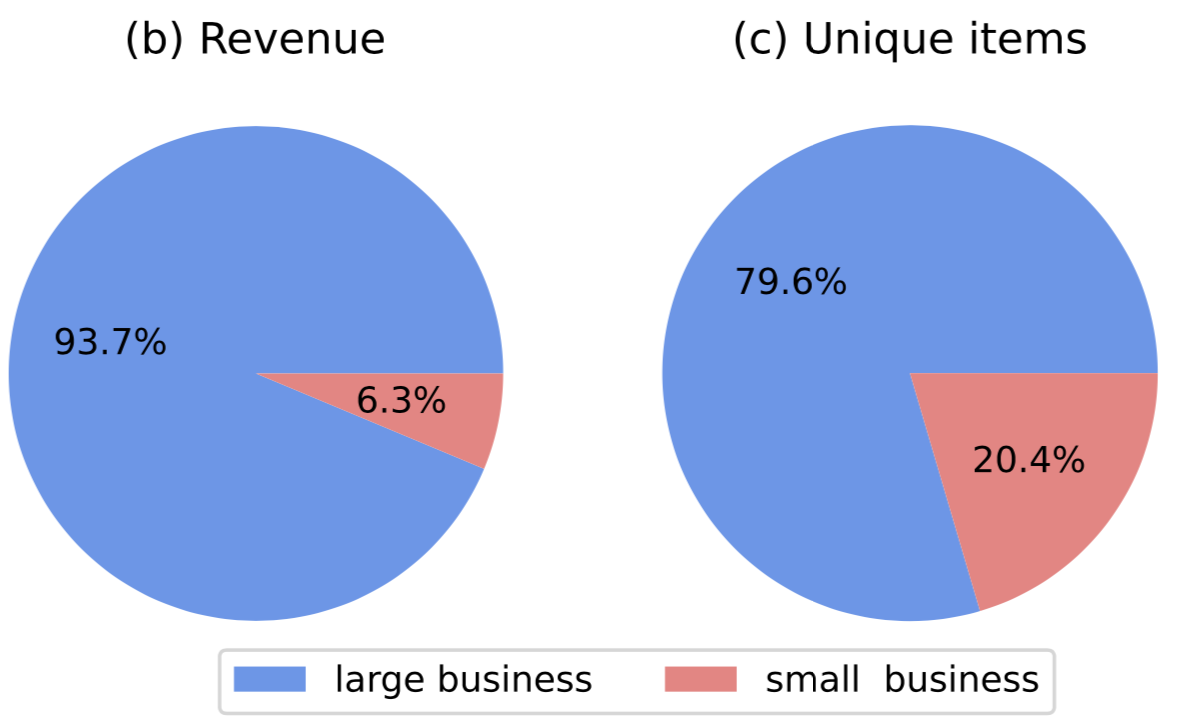}
\end{minipage}
\caption{ Sales data of Genre A from Ichiba platform. After sorting the shops by the number of sales, the top 25\% of shops are categorized as large businesses, and the rest are small businesses.  (a) the histogram of shop's sales. (b) the breakdown of one-year revenue. (c) the breakdown of unique items in large and small businesses. }
    \label{fig:1}
\end{figure}

On the one hand, the small business problem is closely related to the  popularity bias  problem.
Many researchers have explored the popularity bias in recommender systems. Abdollahpouri \etal proposed a regularization-based factorization method to improve long tail recommendations~\cite{abdollahpouri2017controlling} . Brynjolfsson \etal and Park \etal tried to improve  the long-tail recommender system performance via reducing popularity bias and balancing the popularity distributions~\cite{brynjolfsson2006niches,park2008long}. We tried two widely-used solutions, oversampling (Section\ref{sec:nsresult}), and regularization (Section~\ref{sec:fair}). However,  they did not improve the recommendation performance for small businesses shop-wise. Thus, item-level bias approaches cannot solve shop-level bias. Our work focuses on shop-level popularity bias instead of item-level bias.





On the other hand, less popular items from small businesses can be treated as cold-start items. 
One possible solution for cold-start items is the content-based models~\cite{10.1145/3397271.3401060,10.1145/336597.336662}. Instead of using transaction history, these models used content features (i.e., side information) such as user profile and item information. Hybrid models that combine content-based models and collaborative filtering have also been popular~\cite{Li15,Chen12}. The main idea is to map the side information and the feedback to separated low-dimensional representations and then use them together to predict the final interaction. However, hybrid methods still suffer from sample selection bias~\cite{ssb} problem since it is inevitable during training ~\cite{10.1145/3331184.3331268}.
Our experimental results in Section~\ref{sec:motivation} also showed that the state-of-the-art cold-start RS~\cite{Razi21} did not solve the shop-level imbalanced performance. 

In many recent works, meta-learning has been used to improve recommendation performance. Notably, model-agnostic meta-learning (MAML)~\cite{finn2017model} is widely used to solve cold-start problems in RS due to its success in few-shot learning~\cite{lee2019melu, metacs,mwuf}. 
In this scenario, the recommendation for a user is  a task. Each user has their local model  that is updated by their past transaction. After all local models are updates, the global model is updated by minimizing the loss over all training tasks. Then, the learned global model initializes local models for the next update.  With this, the knowledge from local models with sufficient training samples can be transferred to models with less sufficient models and help cold-start users. 


To improve the advertising performance for small businesses, we propose \textit{Meta-Shop}. Meta-Shop leverages the  MAML and builds a model for the shop-level recommendation.
By applying the MAML framework to shop-level tasks, Meta-Shop could solve the popularity bias problem by integrating and transferring knowledge between large and small shops, consequently providing better recommendation performance for small shops. In addition, the learning-to-learn property of MAML enables our model to adapt to cold-start or new shops quickly.
While the recent work MeLU~\cite{lee2019melu}  focused on cold-start user preference estimation, Meta-Shop learns user and item representations for different shops. 
Another advantage of Meta-Shop is its training stability.
MAML is known to have issues such as training instability and slow convergence~\cite{antoniou2018how}. Previous work 
 added additional task-specific parameters or gradient-skip connections to reduce model instability~\cite{Dong2020mamo,DBLP:journals/corr/abs-2010-09291}. By training per shop-level, Meta-Shop reduces the number of tasks without any hyper-parameter setting and thus provides stable training with better generalization.

To the best of our knowledge, this is the first work that considers popularity bias in businesses, learns meta user/item features at the shop-level,
and builds a quick-adaptive shop-level recommender system. Our contributions are as follows:
\begin{itemize}
    \item We study a novel problem on real-world E-commerce platforms, namely improving advertising performance for small businesses (i.e., shops with a few sales history).
    \item We propose Meta-Shop to solve the above problem. Meta-Shop is a  meta-learning-based recommender system.
    It trains a user-item interaction prediction model that quickly adapts to different shops. 
  Instead of emphasizing learning on the large businesses/shops, it transfers knowledge learned from large shops to small shops and improves small shops' prospective user targeting. Also, it can quickly adapt to unseen shops without the past transaction.
    \item Experimental results on a real-world E-commerce dataset and the Movielens1M dataset demonstrate that our method outperforms the  existing production baseline and the state-of-the-art models. Notably, for the recommendation for unseen shops, we achieved  16.6\% relative improvement of Recall@1Million on  Ichiba E-commerce dataset and 40.4\% relative improvement of nDCG@3 on the Movielens dataset.
\end{itemize}

\section{Related Work}
\subsection{Popularity Bias Problem in RS}
The popularity bias problem in RS, also known as the long-tail item popularity problem, has been explored. The researches are mainly fallen into two categories: 1) to improve the performance of RS in the presence of long-tail items; 2) to increase the chance of long-tail items to be recommended.
First of all, researchers worked on improving the overall recommendation performance~\cite{brynjolfsson2006niches}. For example, \citeauthor{park2008long} proposed  splitting tail items from the whole item sets and building a RS based on head items and tail items separately.   
Furthermore, other researches focused on controlling popularity bias to recommend more long-tail items~\cite{abdollahpouri2019managing}. More specifically, \citeauthor{abdollahpouri2017controlling} introduced a regularization-based framework and showed that the regularization could trade-off between recommendation accuracy and coverage of long-tail items.
Zhang  {\em{et al.}}~\cite{Zhang2021mirec} combined curriculum learning and meta-learning to improve the performance of long-tail item recommendations. However, this model requires information about all items popularities in training which can be hard to obtain in real-world cases.

\subsection{Hybrid RS for Cold-start Problems}
Various hybrid methods have been proposed to tackle the cold-start problem, such as collaborative topic regression~\cite{Wang11} and its variants \cite{Gopalan15,Wang15b}. Based on the availability of feedback information, collaborative topic regression interpolated between content-based representations generated from side information by Latent Dirichlet Allocation (LDA)~\cite{Blei03} and feedback-based ones generated from interaction matrices by weighted matrix factorization (WMF). 
 
Recent hybrid methods utilized neural networks to learn representations from side information. To solve the cold-start problem in the music recommendation system, DeepMusic~\cite{Oord13} created multiple embeddings based on contents (e.g., artist biographies and audio spectrogram) using neural networks. Other researchers utilized various autoencoders to learn the representation of side information~\cite{Dong17,Zhang17b,Li18}. 


Some works, such as Dropoutnet~\cite{Volkovs17},  focused on integrating side information and the preference/feedback information. The values of preference representations are set to zero for cold-start items when past interactions are missing.
Bianchi {\em{et al.}} utilized feedbacks from multi-brand retailers and aligned item embeddings across shops using translation models~\cite{fantasticembeddings}. CB2CF~\cite{Barkan19} learned a mapping function from side information to  representation. 
Attention mechanisms were also introduced to improve the recommendation performance. 
ACCM~\cite{Shi18} paid more attention to the cold-start item's side information than other interaction information to make predictions. Raziperchikolaei {\em{et al.}} proposed a hybrid method that creates user representation using side information of items purchased~\cite{Razi21}. With this, this method recommended cold-start items only with side information.

\subsection{Meta-learning}
 Meta-learning, also called learning-to-learn, is thriving as it trains a meta-model that can quickly adapt to new tasks~\cite{10.5555/3305381.3305498,metasurvey}. It has been applied to different areas such as computer vision~\cite{matchingnetwork,antoniou2018how} and natural language processing~\cite{languagesurvey,languagemeta}. 
 Previous researchers learned novel metric functions~\cite{snell2017prototypical,sung2018learning} or designed architectures that rapidly generalize to new tasks~\cite{munkhdalai2017meta,santoro2016meta}.
 Recently, optimization-based methods have achieved massive success in the meta-learning field~\cite{finn2017model}. It can quickly adapt to new tasks with a few samples. %
 
 The success of optimization-based meta-learning in few-shot settings has shed light on the cold-start recommendation systems. Lee \etal and Vartak \etal
 applied meta-learning to RS and treated each user's recommendation as separated tasks~\cite{NIPS2017_51e6d6e6, lee2019melu}. Each task predicts whether a user likes an item or not. Scenario-specific sequential meta learner~\cite{10.1145/3292500.3330726} learned user's behaviors from different scenarios such as ``what to take when traveling'' and ``how to dress up yourself on a party''. Here, tasks are based on scenarios. 
Pan \etal  proposed a two-step model to improve click-through rate predictions for cold-start advertisements (ads) by first training a traditional classification model using warm-start ads and then adding a meta-learning module to fine-tune cold-start item feature embeddings~\cite{10.1145/3331184.3331268}. Besides using meta-learning to learn user and item features, several works also focused on using meta-learning to optimize model structures~\cite{10.1145/3366423.3379999,10.1145/3397271.3401167}.
Dong {\em{et al.}} \cite{Dong2020mamo} introduced memory-augmented meta-optimization (MAMO) for recommender systems. Introducing two memory tensors that store user preference and prediction weight helped the model with better generalization. However, MAMO requires a hyper-parameter to decide the number of user preference groups.

 

\section{Popularity bias for small businesses}
\label{sec:motivation}
This section presents the analysis result showing that existing cold-start item recommender systems suffer from popularity bias issues.

\begin{figure}
    \centering
    \includegraphics[width=6cm]{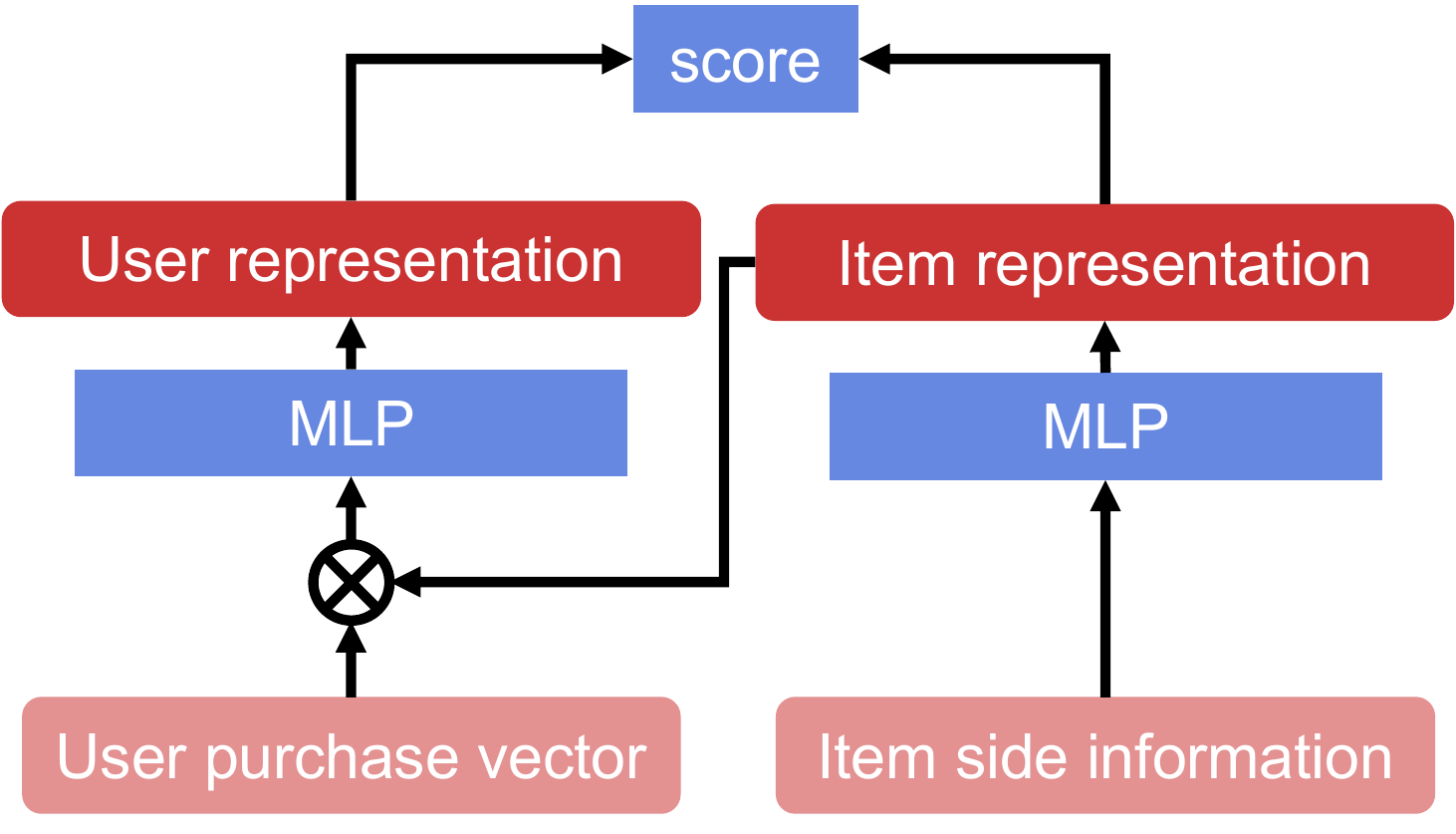}
    \caption{Production baseline model for cold-start item recommendation. $\otimes$ operation is a summation of all item representations of which the user purchased.}
    \label{fig:rocket}
\end{figure}

We analyzed the shop-level recommendation performance using Ichiba's production RS baseline. 
The architecture of the production baseline is shown in Figure \ref{fig:rocket}. It creates item representations from side information such as price and title, then creates user representation by aggregating representations of items the user purchased in the training period. Euclidean distance between the user and item representation is the final interaction score. It utilizes contrastive loss~\cite{1467314} with positive and negative samples to make user and item representations similar when there is interaction and dissimilar when not\footnote{The performance of the baseline model with Mean-squared-error loss was worse than with contrastive loss.}.  

The objective function is as follows:
\begin{align*}
\min_{\theta} &\sum_{(i, u) \in \mathcal{D}^+}\| f_u(N_{u}) - f_i(x_i))\| \\&+ \lambda \sum_{(i,u) \in \mathcal{D}^-} \max(0, m_d - \| f_u(N_{u}) - f_i(x_i))\|)\\
subj&ect\;to\;\; f_u(N_{u}) = \frac{1}{|N_u|}\sum_{i \in N_{u}} f_i(x_i)
\end{align*}
where $\mathcal{D}^+$ are positive samples,  $\mathcal{D}^-$ are negative samples, $N_u$ is a collection of  items purchased by user $u$, $x_i$ is item $i$'s side information, $m_d$ is the margin.


\begin{figure}
    \centering
    \includegraphics[width=7.5cm]{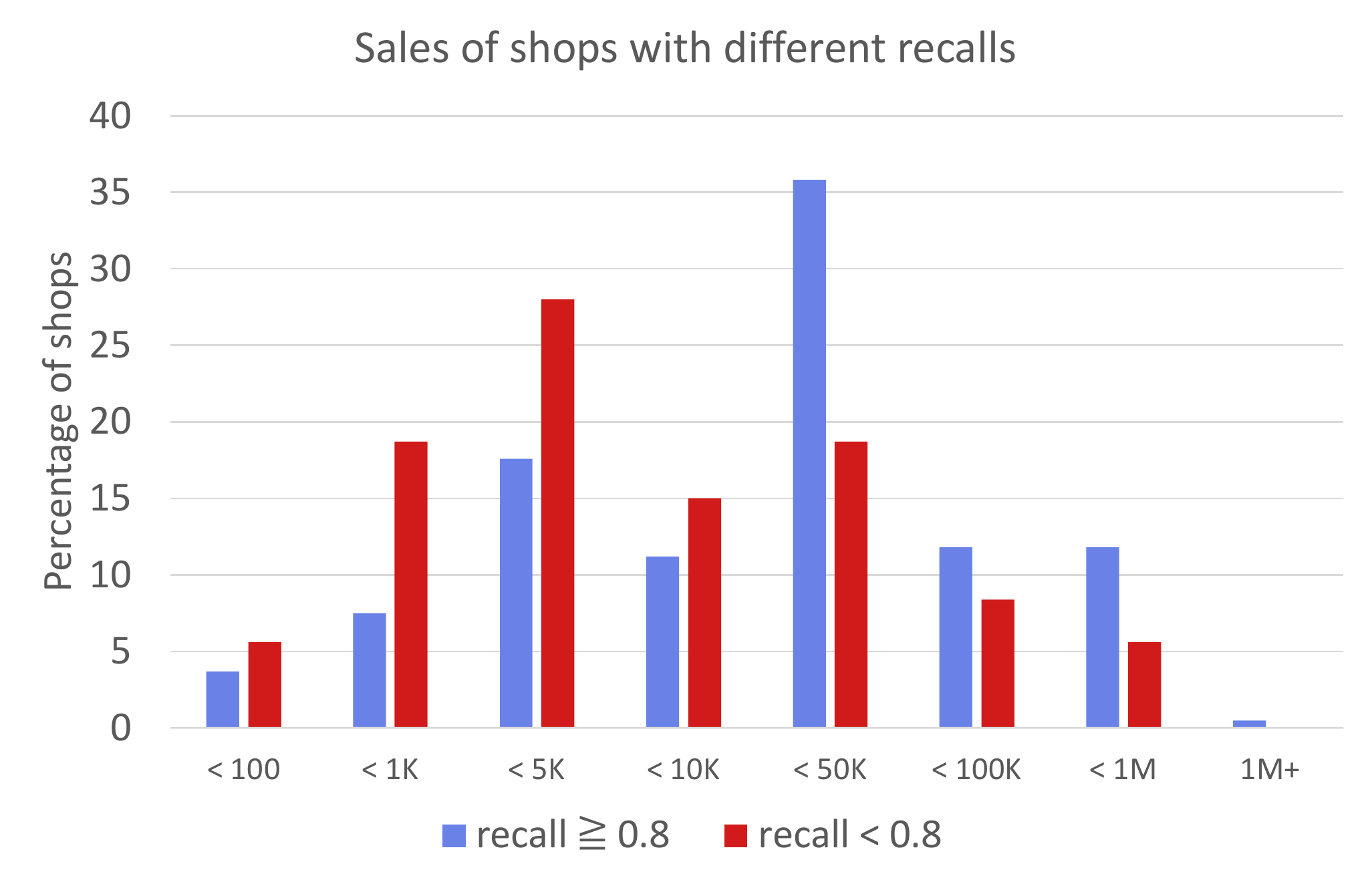}
    \caption{Distribution of the number of sales per test shop in training data. Shops with lower recall have few sales.}
    \label{fig:purchasecount}
\end{figure}

We trained the baseline model using a randomly selected $\sim$30M purchase history with $\sim$3M users and $\sim$1.3M items from purchase data in Ichiba Genre A.
We evaluated the Recall@1M per test shop. The shop recall was computed by averaging Recall@1M of each test item sold by the shop. Based on shop recall, we divided test shops into two groups: shops with Recall@1M $\geq$ 0.8 and shops with Recall@1M $<$ 0.8. 

After computing statistics such as ``number of sales per shop", ``average item price per shop",  and ``average length of item description per shop", we found that the ``number of sales per shop" is most likely a cause for the performance differences. Specifically, bad-performed shops have fewer sales. For example, $59.9\%$ shops with Recall@1M $\geq0.8$ have more than 10k sales in training, while more than $52\%$ shops  with Recall@1M $<0.8$  have less than 5k sales. The details are shown in Figure~\ref{fig:purchasecount}.  
This analysis concluded that \textit{the existing cold-start recommendation models are ineffective for small shops/businesses because of the popularity bias.}

\section{Proposed method}

\label{sec:meta}
In this section, we propose a novel meta-learning framework to solve cold-start item recommendations for small businesses: \textbf{Meta-Shop}. In the Meta-Shop model, 
we define per shop recommendations as separated tasks. 

\subsection{The \textit{Meta-Shop} Model}
\begin{figure}[!t]
    \centering
    \includegraphics[width= 5cm]{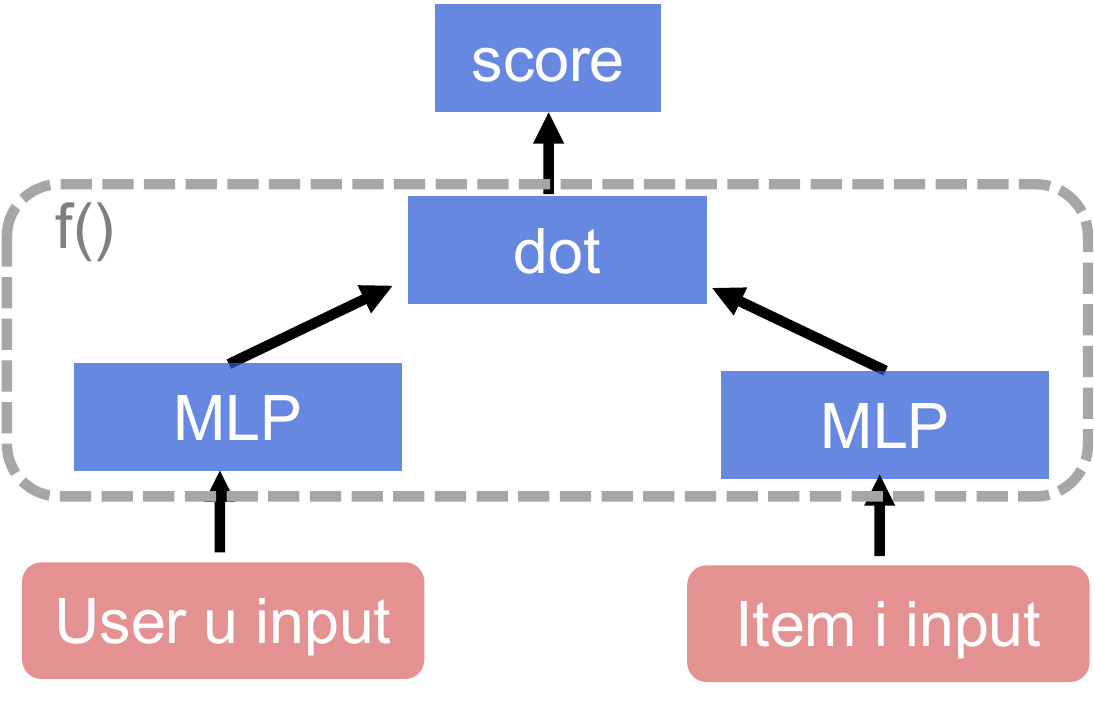}
    \caption{Meta-Shop model: User and item have separated mapping models. The two hidden features are combined by a dot product at the last layer.  Parameters in f() are meta-parameters shared among all shops. }
    \label{fig:metashop}
\end{figure}

Model inputs are user feature $e_u$ and item feature $e_i$. They can be trainable one-hot embeddings or pre-trained representations. 
The model parameter is $\theta$. For shop $p \in \mathcal{T}_{shop}$, we define its loss as  $$\mathcal{L}(\mathcal{D}^p;\theta) = \frac{1}{n_p}\sum_{(i,u, y_{i, u}) \in \mathcal{D}^p}\mathcal{L}(y_{i, u},  \Tilde{y}_{i, u}),$$ where $\mathcal{D}^p$ are shop $p$'s transaction data, i.e., $\{(i, u, y_{i,u}))\}$ (with size $n_p= |\mathcal{D}^p|$). 
$y_{i, u}\in \{0,1\}$ describes a user-item purchase interaction, i.e., $y_{i, u}$ is 1 if user $u$ purchased item $i$ (in shop $p$) and 0 otherwise,  $\hat{y} = f(e_u,e_i; \theta)$ is the predicted score, $f(\cdot)$ is the prediction function and $\theta$ are model parameters. $\mathcal{L}(\cdot, \cdot)$ is loss function, which can be either squared loss: $\mathcal{L}(y_{i, u}, \Tilde{y}_{i, u} ) = (y_{i, u} - \Tilde{y}_{i, u} )^2$, or cross-entropy loss $\mathcal{L}(y_{i, u}, \Tilde{y}_{i, u} ) = y_{i, u} log( \Tilde{y}_{i, u} ) + (1-y_{i, u} )log( 1- \Tilde{y}_{i, u} )$. In this paper, we use squared loss for all experiments. For prediction function $f(\cdot)$, we design the network architecture: \textbf{Meta-Shop} (\textbf{MeSh}), as shown in Figure~\ref{fig:metashop}. 


In MeSh, user and item inputs are separately fed into two different multi-layer perceptrons (MLPs): $h_u = f_u(e_u; \theta_u), h_i = f_i(e_i; \theta_i)$, and at the last layer, the two hidden features are combined by a dot product: $\hat{y} = \langle h_u, h_i\rangle$,
where user preference representation $f_u(e_u; \theta_u)$ and item preference representation $f_i(e_i; \theta_i)$ are:
\begin{align*}
f_u(e_u; \theta_u) &= \sigma( ... \sigma(\sigma( e_u; W_1^u, b_1^u); W_2^u, b_2^u)... ; W_L^u, b_L^u)\\
f_i(e_i; \theta_i) &= \sigma( ... \sigma(\sigma( e_i; W_1^i, b_1^i); W_2^i, b_2^i)... ; W_L^i, b_L^i)
\end{align*}
and $\sigma()$ is an activation function, $W_l^u, b_l^u$ and $W_l^i, b_l^i$are the weight matrix and bias vector of $l$-th layer of MLP $f_u(\cdot; \theta_u)$ and $f_i(\cdot; \theta_i)$ respectively.

\begin{algorithm}[!b]
\SetAlgoLined
\KwResult{ $\theta$}
 Initialization of  $\theta$, stepsize $\alpha, \beta$, number of local updates $K$\;
 \While{not converge}{
  Sample batch of shops $\mathcal{P}$ from $\mathcal{T}_{shop}$\;
  \For{ shop p in $\mathcal{P}$}{
    Set $\theta^p = \theta$\;
     
    Sample $\mathcal{D}^p_{S}$ and $\mathcal{D}^p_{Q}$ from $\mathcal{D}^p$\;
    \For{ k in 1, ..., K}{
    $\theta^p \leftarrow \theta^p-\alpha \nabla_{\theta^p} \mathcal{L}$($\mathcal{D}^p_{S}$; $\theta^p$) \;}
    }
   
global update $\theta$: $\theta \leftarrow \theta -\beta \sum_{p \in \mathcal{P}} \nabla_\theta$ $\mathcal{L}$($\mathcal{D}^p_{Q}$; $\theta^p$)\;
 }
\caption{Meta-Shop Training}
 \label{alg:0}
\end{algorithm}

\begin{algorithm}[]
\SetAlgoLined
\KwResult{ \{$\theta^p$ ; $p \in \mathcal{P}_{inf}$\}}
  Targeted small shops $\mathcal{P}_{inf}$ and their observed transaction data \{$\mathcal{D}^p$ ; $p \in \mathcal{P}_{inf}$\}, number of local updates $K$\;
  Trained meta model parameter  $\theta$, stepsize $\alpha$\;
  \For{ shop p in $\mathcal{P}_{inf}$}{
    Set $\theta^p = \theta$\;
    Sample $\mathcal{D}^p_{S}$ from $\mathcal{D}^p$\;
        \For{ k in 1, ..., K}{
    $\theta^p \leftarrow \theta^p-\alpha \nabla_{\theta^p} \mathcal{L}$($\mathcal{D}^p_{S}$; $\theta^p$) \;}
  }
\caption{Meta-Shop Inference}
 \label{alg:1}
\end{algorithm}

\subsection{Optimization}
We adopt the MAML algorithm~\cite{10.5555/3305381.3305498} and show training steps in Algorithm~\ref{alg:0} and inference steps in Algorithm \ref{alg:1}. In Algorithm~\ref{alg:0},
$\mathcal{T}_{shop}$ is a collection of all shops, which can be either large shops or small shops. For each shop $p$, support dataset $\mathcal{D}^p_{S}$ and and query dataset $\mathcal{D}^p_{Q}$ are i.i.d drawn from $\mathcal{D}^p$, where $\mathcal{D}^p_{S}\cap \mathcal{D}^p_{Q} = \emptyset$. The gradient update for shop $p$ in the inner loop is as follows: 
\begin{align*}
    \theta^p \leftarrow \theta-\alpha \nabla_{\theta} \mathcal{L}(\mathcal{D}^p_{S}; \theta),
\end{align*}
which update one or more gradient steps for meta model parameter $\theta$ for shop $p$. It is important to take a note that the meta model parameter $\theta$ is expected to be general enough for all shops, so that the one step of local gradient update for shop $p$ can achieve a good performance. To attain such a goal, we adopt the standard gradient-based meta-learning framework, for which the meta-objective is formalized as: 
\begin{align*}
 \min_{\theta} \sum_{p \in \mathcal{T}_{shop}} \mathcal{L}(\mathcal{D}^p_{Q};\theta^p)
\end{align*}
In above meta-objective, it is optimized over meta parameter $\theta$, while the value of the objective function is calculated using the parameter $\theta^p$ of each shop $p$. We solve the meta-objective  by MAML method but any other gradient-descent based optimization methods, such as Reptile\cite{nichol2018first},  can also be used here. The global update step for meta parameter $\theta$ can be described as follows: 
\begin{align*}
    \theta &\leftarrow \theta -\beta \sum_{p \in \mathcal{P}} \nabla_\theta \mathcal{L}(\mathcal{D}^p_{Q}; \theta^p)\\
    &= \theta -\beta \sum_{p \in \mathcal{P}} \nabla_\theta \mathcal{L}(\mathcal{D}^p_{Q};\theta-\alpha \nabla_{\theta} \mathcal{L}(\mathcal{D}^p_{S};\theta))\\
    &= \theta -\beta \sum_{p \in \mathcal{P}} \nabla_{\theta'} \mathcal{L}(\mathcal{D}^p_{Q};\theta')|_{\theta' = \theta^p} (I-\alpha\nabla_{\theta''}^2 \mathcal{L}(\mathcal{D}^p_{S};\theta''))|_{\theta'' = \theta}
\end{align*}
where $\nabla \mathcal{L}$ is the first derivative of $\mathcal{L}$, $\nabla^2 \mathcal{L}$ is the second derivative of $\mathcal{L}$ and $I$ is identity matrix. For simplicity, we ignore the second derivative part in application.

\section{Experiments}
In this section, we detail experiments  to demonstrate the effectiveness of our proposed method. We verify its effectiveness with the following questions: (Q1) how good is the performance for all item-level recommendations? (Q2) how good is the shop-level recommendation performance, especially the small businesses'?  The first question considers item-level performance, and the second one  considers shop-level performance. In the end, we perform additional experiments to show other solutions for promoting small businesses, such as changing training data and adding a bias regularizer.
\subsection{Dataset}
We use two datasets in our experiments: Ichiba and Movielens1M~\cite{10.1145/2827872}.  For the Ichiba dataset, we extracted sales history from genre A in Ichiba platform. We used the sales data from January 2020 to September 2020 for training and used October 2020 as test data. We kept shops with at least 13 sales in both training and test time, as the model  requires support and query sets for each shop. 10 sales were randomly selected as the support set in each shop, , and the rest went into the query set. Also, we classified test shops into \textit{new shops} and \textit{existing shops}. New shops are shops never shown in the training period while existing shops exist in the training period. We further separated existing shops into small and large based on the number of sales in training. Specifically, the top 25\% of shops with the most  sales are large shops; the rest are small shops.

For Movielens, we followed MeLU~\cite{lee2019melu}: movies released
before 1998 and after 1998 are training and test sets, respectively. We regard each genre (e.g. ``action'', ``drama", and ``crime.")  as a shop for the experiment, based on the observation that the number of ratings was also biased  on genres.  Among 18 genres, we manually removed 3 genres from the training. They were used as \textit{new shops} in the test.  
Like the Ichiba dataset, we kept at least 13 movie ratings in both training and test for each genre. We will use shops instead of genres for unified notation purposes in the following section. 

Note that all test items did not appear in both datasets in the training period, i.e., all test items are completely cold-start.  
The statistics of the datasets are listed in Table~\ref{tab:dataset}.

\subsection{Models}
\label{sec:models}
For the Ichiba dataset, we compared the proposed method with the following models:
\begin{itemize}
    \item \textbf{LV2}: For each query item, we return the most frequent buyers who purchased items from the same level 2 genre as the query item. Note that genre taxonomy has several levels, from broad genres (lv1, e.g., shoes) to specific genres (lv5, e.g., running shoes). 
    \item \textbf{Baseline}: The production baseline is introduced in Section~\ref{sec:motivation}. Item inputs are learnable one-hot embeddings from side information such as price and title. User inputs are summations of all item representations of which they purchased in the training period. 
    \item \textbf{Meta-Shop-interaction} (\textbf{MeSh-i}): Instead of training user and item embeddings separately, in MeSh-i, user and item inputs are concatenated ($x_0 = [e_u; e_i]$) before feeding into a MLP network. Output is $ 
\hat{y} = f(x_0; \theta)$, where $f(x_0;\theta) = \sigma( ... \sigma(\sigma( x_0; W_1, b_1); W_2, b_2)... ; W_L, b_L)$ and $\sigma()$ is an activation function, $W_l, b_l$ are the weight matrix and bias vector of $l$-th layer of MLP $f(\cdot; \theta)$

    \item \textbf{MeLU}~\cite{lee2019melu}: MeLU is the state-of-the-art meta-learning framework for cold-start recommendation. It applied model-agnostic meta-learning (MAML) for individual users to learn a user preference estimator. Since in our setting, our goal is prospective user targeting--recommend users to sellers, which is the opposite of user preference estimation, we trained MeLU for individual items. 
    The model structure is the same as MeSh-i.
\end{itemize}

For the Movielens dataset, we followed MeLU~\cite{lee2019melu} and compared our model with \textbf{Wide \& Deep}~\cite{10.1145/2988450.2988454} and \textbf{MeLU}. Wide \& Deep shared the same model structure as MeLU. The difference is that Wide \& Deep used a standard SGD optimizer without MAML-style optimization.

\begin{table}[!t]
    \centering
   \scalebox{0.7}{
    \begin{tabular}{c|c|c|c|c|c}
            \toprule
        \textbf{Ichiba} & \textbf{Use}r& \textbf{Item} & \textbf{New shop} & \textbf{Existing shop}& \textbf{Purchases} \\ 
        \midrule
        Train& 9,095,846 & 1,409,483& 0& 7,035& 40,118,920\\
        Test & 9,387,803& 894&  9  & 354& 23,325 \\
    \toprule
        \textbf{Movielens} & \textbf{User}& \textbf{Item} & \textbf{New genre} & \textbf{Existing genre}& \textbf{Ratings} \\ 
       \midrule
        Train & 5943 & 2526 & 0 & 15 & 146,023\\
        Test & 5847 & 621& 3 & 15 &126,064 \\

\bottomrule
\end{tabular}
}
   \caption{Dataset statistics.}
    \label{tab:dataset}
\end{table}

\begin{figure*}[t]
    \centering
    \includegraphics[width=15cm]{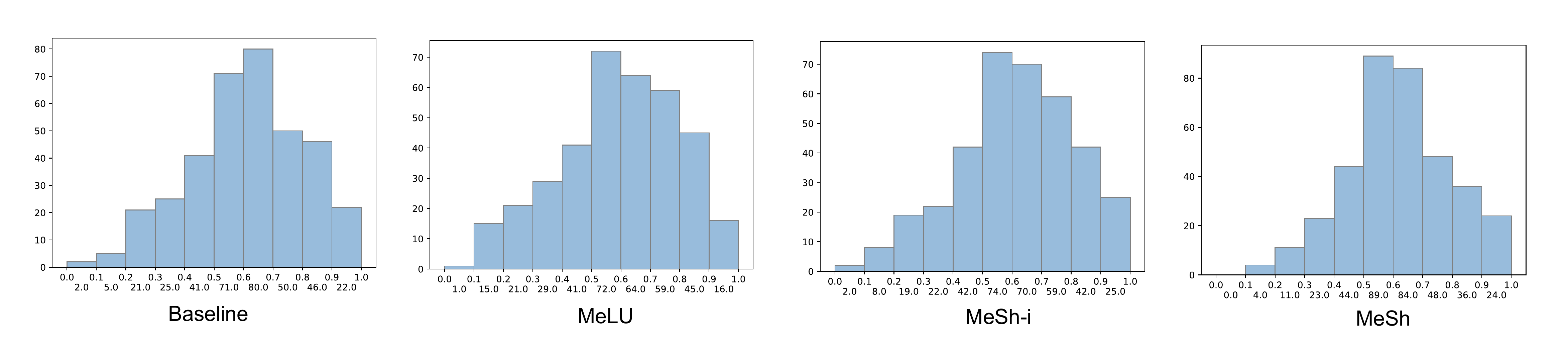}
    \caption{Distribution of test shops with different Recall@1M. The distribution for MeSh-i and MeSh are more right-skewed than the ones for Baseline and MeLU, which means our proposed methods tend to give overall better performance in the shop-level recall.}
    \label{fig:metashop-hist}
\end{figure*}
\subsection{Evaluation Metrics}
For the Ichiba dataset, we computed recall (R). For Movielens, we computed  mean absolute error (MAE) and normalized discounted cumulative gain (nDCG). The detailed metrics are:
\begin{align*}
    R@k = \frac{1}{n_t}\sum_{p\in \mathcal{T}_{shop}}{\frac{1}{n_p}\sum_{i,u \in \mathcal{D}^p}{\frac{R^p_k}{k}}}
\end{align*}

\begin{align*}
    MAE = \frac{1}{n_t}\sum_{p\in \mathcal{T}_{shop}}{\frac{1}{n_p}\sum_{i,u \in \mathcal{D}^p}{|y_{i,u}-\hat{y}_{i,u}|}}
\end{align*}
\begin{align*}
    nDCG_k = \frac{1}{n_t}\sum_{p\in \mathcal{T}_{shop}}{\frac{DCG_k^p}{IDCG_k^p}}; DCG_k^p = \sum_{r=1}^k\frac{2^{y_{pr}}-1}{log_2(1+r)}
\end{align*}
where $n_t$,$n_p$, $R_k^p$ are the number of shops in all shops $\mathcal{T}$, number of transactions in shop $p$ and the number of true purchases in shop $p$ from the top-k ranked predicted purchases. $y_{pr}/\hat{y}_{pr}$ and $IDCG_k^p$ are  the real/predicted rating of user i for the r-th
ranked item and the best possible $DCG_k^p$ for shop $p$, respectively. 
We used different evaluations for the two datasets because the Ichiba dataset has implicit feedback (purchase data) while Movielens has explicit feedback (1-5 ratings).

For all evaluation metrics, we have \textit{item-level} and \textit{shop-level} results. Take the Ichiba dataset evaluation as an example:
Assume total $N$ items, $M$ shops, $r_i$ is the $i$-th item recall value, $s_p$ is the $p$-th shop recall value,  $\mathcal{D}^p$ is a set of items in shop $p$, and $n_p = |\mathcal{D}^p|$ is the number of items in shop $p$, then item-level and shop-level recall are computed as: $R_{item} = (\sum_{i=1}^N r_i)/N$, $R_{shop} = ( \sum_{p=1}^M  s_p)/M$, where $s_p = (\sum_{i \in \{\mathcal{D}_p\}} r_i)/n_p $. 

\subsection{Implementation Details}
For the Ichiba dataset, the user and item input features are pretrained $4096$-dim embeddings from the baseline we described in Section~\ref{sec:motivation}. For the Movelens dataset, we followed MeLU's implementation\footnote{https://github.com/hoyeoplee/MeLU} and performed an embedding process for the inputs. Specifically, we used categorical information: user's gender, age, occupation, area, and item's rate, genre, director, and actor. Each category embedding's dimension is $32$. User and items are represented using the concatenated embeddings of their information. 

In all neural network-based models, the MLP network is a two-layer MLP with first and second dimensions 128 and 100.
We performed hyper-parameter tuning and selected the best stepsizes $\alpha$: $5e^{-6}$ and $\beta$: $5e^{-5}$.  
The number of local updates K per shop is set to $2$. 
All experiments were performed on an NVIDIA GeForce GTX 1080 Ti. All codes were written with Tensorflow 2.1.0.

\begin{table}[!t]
    \centering
 \scalebox{0.8}{ 
\begin{tabular}{l|l|c|c||c|c}
       \toprule
   \multirow{2}{*}{  Type }&  \multirow{2}{*}{  Method}& \multicolumn{2}{c||}{ Item-level} &   \multicolumn{2}{c}{Shop-level} \\
   \cline{3-6}
    & & R@500K & R@1M & R@500K & R@1M\\
   \hline 
   \multirow{5}{*}{ \textbf{New Shop}}&
   LV2 &  0.223  & 0.331&  0.067& 0.112 \\
  & Baseline   & 0.329 & 0.521& 0.306 &0.482 \\
      & MeLU  & 0.349 & 0.450& 0.302&0.394\\
       &MeSh-i &   \textbf{0.361} &   0.523 & 0.324&0.451\\
     &  MeSh & 0.326  & \textbf{0.530} &\textbf{0.331} &\textbf{0.562}\\
       
      \hline
    \multirow{5}{*}{\shortstack[l]{\textbf{Existing Shop}\\  \textbf{(Small)}}} &
       LV2  &   0.239 	&0.322  &0.239	&0.327 \\  
       & Baseline  &    0.384	& 0.538&0.370 &0.529 \\
       & MeLU &  0.388 &	0.520 &0.391 &0.520 \\
       & MeSh-i  & \textbf{ 0.396} & \textbf{0.551}	& \textbf{0.394}&0.546 \\
       & MeSh  &  0.385	&0.541 &0.377	& \textbf{0.559}\\
       
       \hline
           \multirow{5}{*}{\shortstack[l]{\textbf{Existing Shop}\\  \textbf{(Large)}} } &
       LV2   &   0.391 &   0.494 &0.337 & 0.448 \\  
       & Baseline &   0.506&   0.662& 0.467 &0.626 \\
       & MeLU  & \textbf{ 0.538}& 0.651 & \textbf{0.492} & 0.607 \\
       & MeSh-i    & 0.517 &  \textbf{0.667} &  0.475&0.625\\
       & MeSh   &  0.475 &  0.652 & 0.444& \textbf{0.627} \\


        \bottomrule
    \end{tabular}
    }
    \caption{ Item-level and shop-level recall  on Ichiba dataset.}
    \label{tab:recall_colditems_ichiba}
\end{table}

\begin{table}
   \scalebox{0.8}{ 
\begin{tabular}{l|l|c|c|c||c|c}
       \toprule
  \multirow{2}{*}{ Type}&  \multirow{2}{*}{ Method}& \multicolumn{3}{c||}{ Item-level} &\multicolumn{2}{c}{Shop-level} \\
  \cline{3-7}
  &&MAE& nDCG1 & nDCG3 & nDCG1 & nDCG3 \\
\hline
   \multirow{4}{*}{ \textbf{New Shop}}&Wide \& Deep & 1.017 &  0.250 & 0.319 & 0.221& 0.292\\
      & MeLU & \textbf{0.954}& 0.237& 0.433 & 0.307&0.391 \\
       &MeSh-i& 0.981& \textbf{0.309}& \textbf{0.453}& \textbf{0.345} & \textbf{0.410 } \\
     &  MeSh& 0.985& 0.288& 0.437 &0.238 & 0.367\\
       
      \hline
\multirow{4}{*}{\textbf{Existing Shop} } &
       Wide \& Deep & 1.001 &  0.268&  0.368 & 0.288 &0.380\\
      & MeLU& \textbf{0.924}& \textbf{0.324} & \textbf{0.433} & 0.321 &0.413\\
      & MeSh-i & 0.968& 0.302& 0.414 & \textbf{0.321}& \textbf{0.416}  \\
      & MeSh & 0.974& 0.296& 0.409 &0.313 & 0.410 \\
        \bottomrule
    \end{tabular}
    }
 \caption{Item-level and shop-level evaluations on Movielens dataset.}
    \label{tab:recall_colditems_ml1m}   
\end{table}

\subsection{Item-level Performance}
\label{sec:item-level}
 We summarize item-level performance results in Tables~\ref{tab:recall_colditems_ichiba} and \ref{tab:recall_colditems_ml1m}. For the Ichiba dataset, MeSh performs the best for the \textit{new shops}. It has a 60.1\% relative improvement compared to LV2 and 17.8\% to MeLU for R@1M. For the \textit{existing shops}, MeLU becomes competitive for the large shops. It is reasonable since MeLU may see more items transactions from large shops
 in each user's support/query set. With more  training samples from large shops, MeLU will have more advantages. 
 For the Movielens dataset, we see similar results, except that MeLU outperforms MeSh in the \textit{existing shop} section. The reason may be that the explicit feedback from Movielens further helps MeLU learn better item representations during item-level optimization. At the same time, the  Ichiba dataset experiment only utilizes 0 or 1 feedback. Also, we notice that MeLU was not the best model for new shops on Movielens, 
 which implies possible overfitting. 

\subsection{Shop-level Performance}
\label{sec:shop-level}
In Table~\ref{tab:recall_shop_ichiba}, we compute the portion of shops with shop-level R@1M above a specific value and mean and variance of shop-level R@1M of each shop. Our model outperforms the rest methods. Specifically, in MeSh, more than 77\% of shops have R@1M more than 0.5. It is twice more than LV2 and 6.9\% more than MeLU. It also achieves a 2.2\% relative improvement of shop-level R@1M  compared to baseline while keeping a smaller variance,\textit{ which indicates that our model is more robust across all different-sized shops.} We plot the distribution of test shops with different R@1M in Figure~\ref{fig:metashop-hist}. \textit{The smooth tail of the distribution matches the fact that MeSh has a smaller shop-level recall variance.}
Even though MeLU performs the best in item-level evaluation on the Movielens dataset, it losses its advantage when  considering shop-level performance, as shown in Table~\ref{tab:recall_shop_ml1m}. In MeSh-i, more than 72\% of shops have a nDCG@3 score of more than 0.35. It also achieves a 1.5\% relative improvement of shop-level nDCG@3 with a smaller variance than MeLU.
In summary, the above results indicate that MeSh and Mesh-i provide  less-biased recommendations for all shops and movie genres.

To further understand the impact of Meta-Shop on small business advertisements, we focus on the \textit{new shop}'s shop-level results. New shops belong to the small business scenario as they do not exist in the training set and only require 10 transactions for the local update before the test. From the right side of Tables \ref{tab:recall_colditems_ichiba} and \ref{tab:recall_colditems_ml1m}, we see Meta-Shop models outperform  others in new shop sections on both datasets.   On the Ichiba  dataset, the relative improvement of MeSh over Baseline for the new shops and large existing shops are 16.6\% and 1.6\% for R@1M. On Movielens, the relative improvement of MeSh-i over Wide \& deep for the new shops and large existing shops are 40.4\% and 9.5\% for nDCG@3.  \textit{More significant improvements for new shops indicate that our model truly helps small business advertisement.}

Table~\ref{tab:gmv} lists the Gross Merchandise Value (GMV) improvement over the LV2 model on the Ichiba dataset. The GMV improvement is more than 200\% in the new shop section when using MeSh which further proves the effectiveness of our model in helping small businesses.

\begin{table}[!t]
    \centering
                \caption{Left: The portion of shops with R@1M above a specific value on the Ichiba  dataset. Right: Shop-level R@1M mean and variance. $\uparrow$ means the metric the higher, the better, $\downarrow$ means the metric the lower, the better.}
    \label{tab:recall_shop_ichiba}
       \scalebox{0.8}{
    \begin{tabular}{l|c|c|c|c||c|c}
       \toprule
        Model & $\geq0.5$& $\geq0.6$& $\geq0.7$ & $\geq0.8$ & Mean$\uparrow$ & Variance $\downarrow$ \\
       \hline
       LV2 & 0.361 & 0.215 & 0.140& 0.066 &0.424& 0.226 \\
        Baseline  & 0.741 & 0.545  &   0.325 & 0.187 & 0.602 & 0.197  \\
        MeLU & 0.705 & 0.507 & 0.331 & 0.168& 0.589 & 0.208\\
        MeSh-i   &0.744 & 0.540 &  0.347   & 0.185 & 0.609 &  0.202\\
        MeSh  &  0.774 & 0.529  & 0.298 &   0.165&\textbf{0.615} & \textbf{0.177}\\
        
        \bottomrule
    \end{tabular}
    }
\end{table}

\begin{table}[!t]
    \centering
            \caption{Left: The portion of shops with nDCG@3 above a specific value on the Movielens dataset. Right: Shop-level nDCG@3 mean and variance.}
    \label{tab:recall_shop_ml1m}
       \scalebox{0.8}{
    \begin{tabular}{l|c|c|c|c||c|c}
       \toprule
       Model & $\geq0.35$& $\geq0.40$& $\geq0.45$ & $\geq0.50$ &  Mean$\uparrow$ &  Variance $\downarrow$ \\
       \hline
       Wide \& Deep & 0.388& 0.333& 0.167& 0.167&0.366& 0.311 \\
       MeLU  &  0.500& 0.389& 0.333& 0.278&0.409&0.112 \\
        MeSh-i   & 0.722& 0.444& 0.389& 0.167& \bf{0.415}& \bf{0.104}\\
        MeSh  &  0.556 & 0.389& 0.333& 0.222& 0.403& 0.110\\
        
        \bottomrule
    \end{tabular}
    }
\end{table}

\begin{table}[]
    \centering
        \caption{GMV improvement over LV2 on Ichiba dataset.}
    \label{tab:gmv}
    \scalebox{0.8}{
    \begin{tabular}{c|c|c|c}
    \toprule
     Method 
         & New Shop & Existing Shop (Small) & Existing Shop (Large) \\
     \hline
     LV2 & - & - & -\\
     Baseline& +104.8\% & +80.4\% & +70.9\% \\
     MeLU &+62.7\% & +65.8\% & +50.1\%\\
     Mesh-i&+101.6\% & +81.4\% & +64.7\%\\
     MeSh & +235.2\% & +70.1\% & +67.3\%\\
     \bottomrule
    \end{tabular}
    }
\end{table}

\textbf{Note:} Between MeSh-i and MeSh, MeSh-i performed better in Movielens experiments, while MeSh performed better in Ichiba  experiments. However, we must  mention that MeSh-i cannot store user/item final representations separately and has to feed in raw input data to compute the score. It is less convenient at inference than MeSh, which can pre-store all representations of users and items and compute the score using the dot product. We see similar concerns in previous work~\cite{10.1145/3292500.3330726}.

\begin{table}[!t]
\centering
    \caption{Performance of one-shop training on the Ichiba dataset.}
    \label{tab:single-ichiba}
\scalebox{0.8}{
    \centering
    \begin{tabular}{l|l|c|c}
    \toprule
    Type & Method &  R@500K & R@1M  \\
    \midrule
      \multirow{2}{*}{ Small}   & One-shop   & 0.096 $\pm$ 0.172 & 0.157 $\pm$ 0.195 \\
      &Mesh &  0.408 $\pm$ 0.153 & 0.641 $\pm$ 0.146\\
       \hline
      \multirow{2}{*}{  Large} &One-shop  &  0.061 $\pm$ 0.052&  0.129 $\pm$  0.132 \\
       &Mesh &  0.471  $\pm$ 0.137& 0.651  $\pm$ 0.137\\
         \bottomrule
    \end{tabular}
    }
\end{table}

\begin{table}[!t]
    \caption{Performance of one-shop training on Movielens dataset.}
    \label{tab:single-ml1m}
\scalebox{0.8}{
    \centering
    \begin{tabular}{l|l|c|c|c}
    \toprule
    Type & Method & MAE&  nDCG1 & nDCG3  \\
    \midrule
      \multirow{2}{*}{ Small}   & One-shop & 1.303 $\pm $0.164 &  0.093 $\pm$ 0.002& 0.126$\pm$ 0.010\\
       &Mesh &  0.890 $\pm $ 0.002 & 0.100 $\pm $0.016 &  0.143 $\pm $ 0.020  \\
       \hline
      \multirow{2}{*}{  Large} &One-shop& 1.316 $\pm$0.104 &  0.092$\pm$ 0.009& 0.119 $\pm$ 0.009 \\
        &Mesh &  0.917$\pm$ 0.004& 0.082  $\pm$ 0.001&  0.114  $\pm$ 0.004\\
         \bottomrule
    \end{tabular}
    }

\end{table}  

\subsection{Additional Experiments}
\subsubsection{Stability of Meta-Shop training}
\label{sec:stability}
To ensure the stability of MAML training, we found the best settings through ablation study---using 2-step local updates, adam optimizer and MSE loss. We show MeSh training loss on the Ichiba dataset in Figure~\ref{fig:1}. It demonstrates the training stability and the reason for selecting MSE instead of other losses.

\begin{figure}
    \centering
    \includegraphics[width=0.45\linewidth]{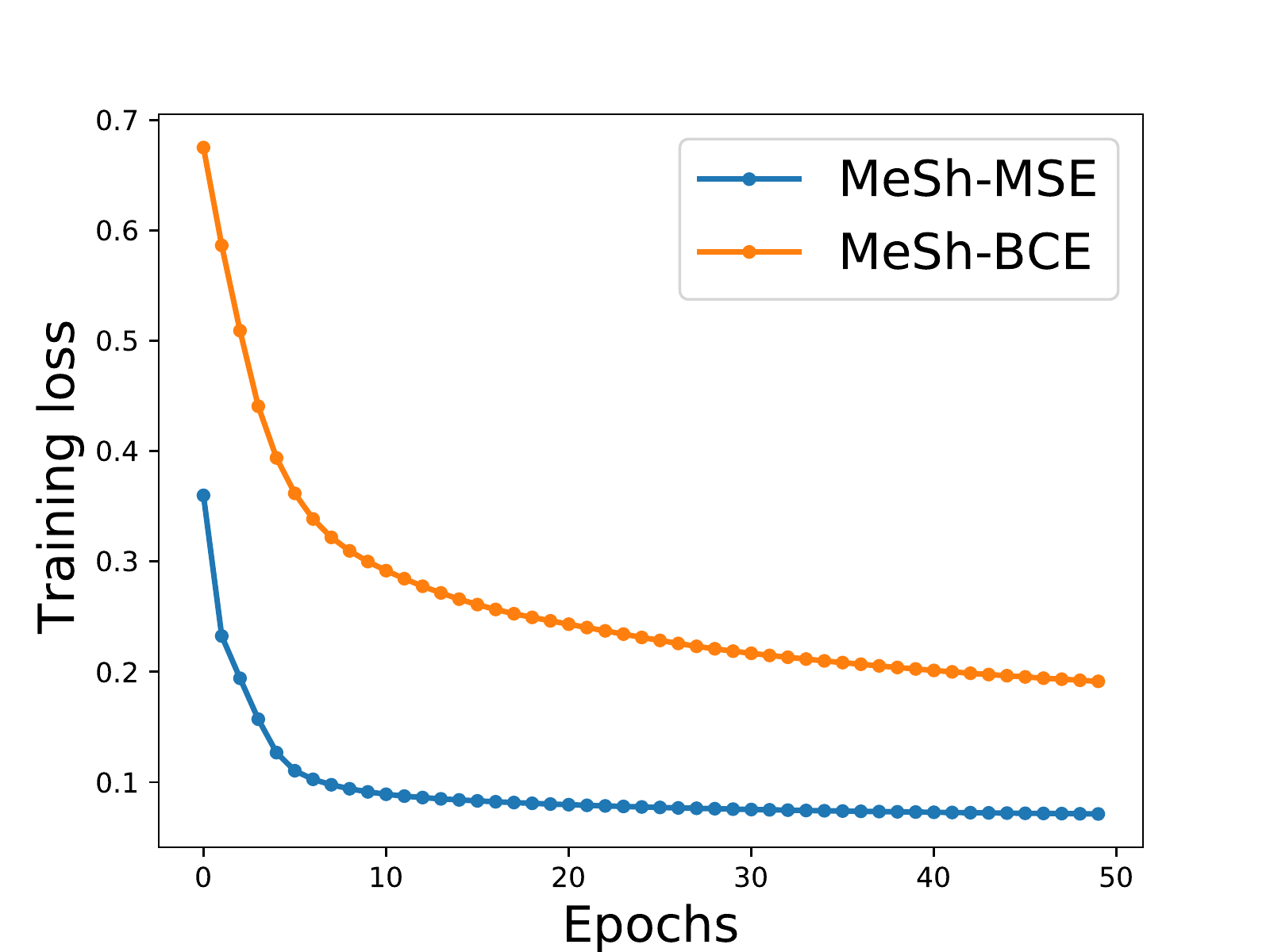}
    \vspace{-0.35cm}
    \caption{Training loss with MSE and BCE losses.}
    \label{fig:1}
\end{figure}

The number of tasks also plays an important role for training stability and performance. Given the Ichiba dataset, we define tasks based on unique items, users and shops; Item-based and shop-based are MeLU and MeSh models defined in Section~\ref{sec:models}. User-based model treats each user's preferences as each tasks as the original MeLU paper proposed. Item-based and user-based models have a total 459K and 1.3M tasks respectively while shop-based only has 7K tasks. We believe smaller number of tasks improves performance more as we show in Table~\ref{tab:difftasks}. Too many tasks will affect model's generalization and convergence speed as shown in ~\cite{antoniou2018how}.

\begin{table}[]
    \centering
        \caption{Item-level R@1M of models with different definitions of tasks on the Ichiba dataset.}
    \label{tab:difftasks}
    \scalebox{0.8}{
    \begin{tabular}{c|c|c|c|c}
    \toprule
  Method 
         & \# of Tasks &New Shop & Existing Shop (S) & Existing Shop (L) \\
     \hline
     Item-based & 459,178 &0.450  & 0.520  &0.651  \\
     User-based&  1,312,017 & 0.519& 0.528  &  0.652\\
     Shop-based & 7,035 &\bf{0.530} & \bf{0.541}  &  \bf{0.652}\\
     \bottomrule
    \end{tabular}
    }
\end{table}

\subsubsection{Meta-Shop versus One-shop training}
\label{sec:ablation}
One practical solution for sample bias in the real world is to train separate models for each shop. Here, we randomly select 6 shops from small and large existing shops respectively, train and test each shop separately using the Wide \& Deep model, namely the one-shop model.  The difference between one-shop and Wide \& Deep is that the former is trained only using one shop's user-item (all available) interaction data  instead of all shops'.
We report the mean and variance of the shop-level test results in Table~\ref{tab:single-ichiba}.  Results on Movielens dataset with 2 shops in each section are in Table~\ref{tab:single-ml1m}. 
One-shop model results are more than 4 times worse than MeSh's on the Ichiba  dataset for R@1M and more than 1.4 times worse on the Movielens for MAE. In summary, one-shop training is not a good solution for sample bias, especially when the dataset is highly sparse and long-tailed like the Ichiba dataset.

\subsubsection{Negative oversampling}
\label{sec:nsresult}
\begin{figure}[!t]
    \centering
    \includegraphics[width=9cm]{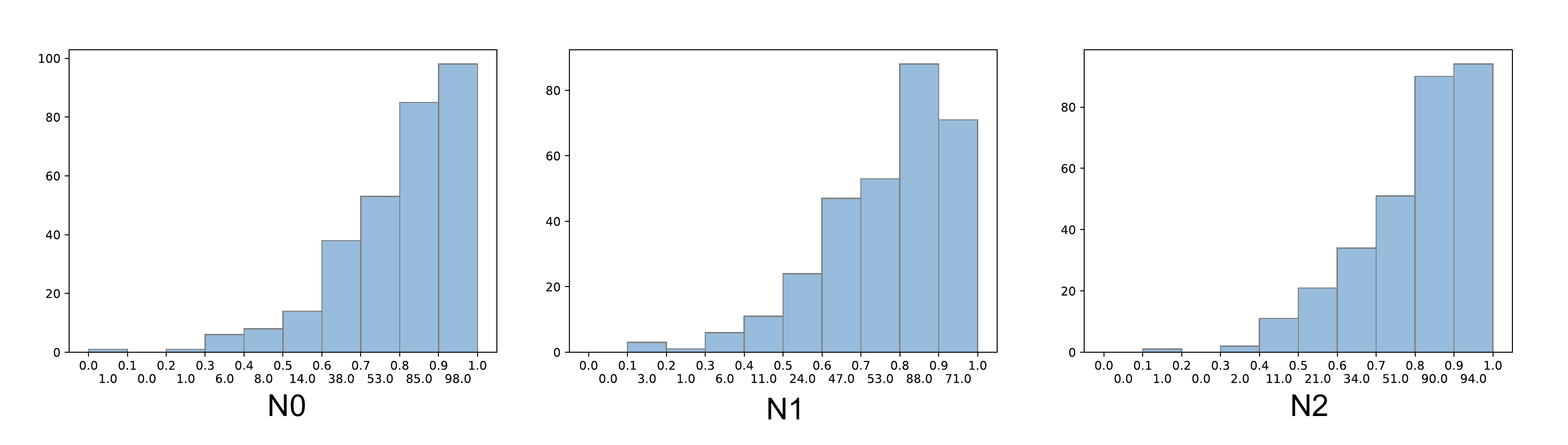}
        \caption{Distribution of test shops with different R@1M. x-axis is the recall value, y-axis is the number of shops, bottom integers are counts of each bar.}
    \label{fig:ns-hist}
\end{figure}

We try to remove sample bias using different negative sampling strategies with the Baseline we described in Section~\ref{sec:motivation}. For each positive item-user purchase pair, we draw one item-user negative pair. Here, we need shop size information during the training. We define
small shops be shops with less than $N$ sales in training. $N$ is set to be the median of the number of sales for all shops in training.  The rest are large shops. We tested three negative sampling methods:
\begin{itemize}
    \item N0: The original NS; the negative user comes from all users who purchased items from different Level 3 genres.
    \item N1: The negative user is selected either from all users who purchased from small shops or users who purchased from different level 3 genres with 0.5 probability each. 
        \item N2: For items in large shops, we use N0, for items in small shops, we use N1.
\end{itemize}
Compared to N0 and N2, N1 has more training users who purchased from small shops. N2 is a combination of N0 and N1. In Table~\ref{tab:recall_ns},  N0 has the highest Recall@1M in both item-level and shop-level evaluation. N0 and N2 keep a similar recall at the shop-level, while N2 achieves a much smaller variance.
In Figure~\ref{fig:ns-hist}, we show the recall distribution of test shops.  The slop gets smaller from N0 to N1, which indicates the recall distributes more evenly among shops. 
  In summary, even though N2 achieves lower shop-wise performance variance, the overall performance was similar to N0. Nevertheless, the selecting negative samples requires more computation costs.

\begin{table}[]
      \centering
              \caption{ R@1M using different negative sampling on Ichiba dataset.}
              \label{tab:recall_ns}
      \scalebox{0.8}{
        \begin{tabular}{c|c||c|c}
        \toprule
       \multirow{2}{*}{ NS} & Item-level& \multicolumn{2}{c}{ Shop-level} \\
        \cline{2-4}
            &   Mean &  Mean &  Variance   \\
       \midrule
        N0 & \textbf{0.838}&\textbf{0.807}&  0.166\\
        N1  & 0.807& 0.769& 0.174\\
        N2  &0.835 & 0.804& \textbf{0.159}\\
        \bottomrule
        \end{tabular}
        }
\end{table}

\subsubsection{ Debiasing regularization }
\label{sec:fair}
We applied a regularizer to the loss function to see if it helps reduce shop-size bias. Followed by the idea in ~\cite{fairmeta}, the regularizer improves the likelihood of items in small shops being recommended. The definition of small/large shops is similar to Section~\ref{sec:nsresult}:
$
\mathcal{L}_{new}(D^p; \theta) = \mathcal{L}(D^p; \theta) + \gamma\mathcal{R}(D_{small}^p;\theta)
$,
where the regularizer 
\begin{align*}
\mathcal{R}(D_{small}; \theta) & = 1 - Prob(\text{shop is recommended }| D_{small}) \\
& \approx 1 - \frac{1}{|D_{small}|} \sum_{(i,u) \in D_{small}} score(i, u)
\end{align*}

Table~\ref{tab:fair} shows experimental results on the Ichiba dataset with different $\gamma$, which is the weight to balance between original loss and the regularizer. Bigger $\gamma$ gives bigger weights on the regularizer. With $\gamma = 0.8$, we see improvement over small existing shop's recommendation. Nevertheless, the new shop's recommendation accuracy was dropped. We also tried bigger $\gamma$; however, it did not return better results. In conclusion, the regularizer only helps small shops that existed in training, but not new shops that did not exist.
\begin{table}[]
    \centering
        \caption{Item-level R@1M of models with different regularization weights on Ichiba dataset.}
    \label{tab:fair}
    \scalebox{0.8}{
    \begin{tabular}{c|c|c|c}
    \toprule
      $\gamma$    & New Shop & Existing Shop (Small) & Existing Shop (Large) \\
     \hline
     0.8 & 0.519& 0.551 & 0.651   \\
     0.01&  0.525 & \textbf{0.549} & 0.648 \\
     0 & \textbf{0.530} &0.541  & \textbf{0.652} \\
     \bottomrule
    \end{tabular}
    }
\end{table}

\section{Conclusion}
In this paper, we proposed Meta-Shop to improve advertising performance for small shops with limited sales. It utilized a meta-learning strategy to learn parameters that can be adapted quickly to unseen shops. The knowledge learned from large shops helped small shops to learn better user and item features. The experimental results showed that Meta-Shop significantly improved recommendation performance for small businesses. 

\bibliographystyle{ACM-Reference-Format}
\bibliography{sample-base}
\end{document}